\documentclass{IEEEtran}

\IEEEoverridecommandlockouts

\usepackage[noadjust]{cite}
\usepackage{amsmath,amssymb,amsfonts}
\usepackage{algorithmic}
\usepackage{graphicx}
\usepackage{textcomp}
\usepackage{xcolor}
\usepackage[binary-units=true]{siunitx}
\usepackage{float}
\usepackage{tabularx}
\usepackage{hhline}
\usepackage{mathtools}
\usepackage{booktabs}
\usepackage[percent]{overpic}
\usepackage{tikz}

\usepackage[acronym]{glossaries}

\usepackage{fancyhdr}

\usepackage[utf8]{inputenc}

\def\BibTeX{{\rm B\kern-.05em{\sc i\kern-.025em b}\kern-.08em
    T\kern-.1667em\lower.7ex\hbox{E}\kern-.125emX}}
    
\definecolor{darkred}{HTML}{A8322C}
\newcommand{\revA}[1]{\textcolor{black}{#1}}
    
\newacronym{rrf}{RRF}{RADAR Repetition Frequency}
\newacronym{qam}{QAM}{Quadrature Amplitude Modulation}
\newacronym{rri}{RRI}{RADAR Repetition Interval}
\newacronym{iot}{IoT}{Internet of Things}
\newacronym{radar}{RADAR}{Radio Detection and Ranging}
\newacronym[plural=TCNs, firstplural={Temporal Convolutional Neural Networks (TCNs)}]{tcn}{TCN}{Temporal Convolutional Neural Networks}
\newacronym{dft}{DFT}{Discrete Fourier Transform}
\newacronym{sot}{SOT}{Signal-over-Time}
\newacronym{sor}{SOR}{Signal-over-Range}
\newacronym{rfdm}{RFDM}{Range Frequency Doppler Map}
\newacronym{hci}{HCI}{Human-computer Interface}
\newacronym{har}{HAR}{Human activity recognition}
\newacronym[plural=HMMs, firstplural={Hidden Markov Models (HMMs)}]{hmm}{HMM}{Hidden Markov Models}
\newacronym[plural=CNNs, firstplural={Convolutional Neural Networks (CNNs)}]{cnn}{CNN}{Convolutional Neural Networks}
\newacronym[plural=ANNs, firstplural={Artificial Neural Networks (ANNs)}]{ann}{ANN}{Artificial Neural Networks}
\newacronym[plural=LSTMs, firstplural={Long Short-Term Memories (LSTMs)}]{lstm}{LSTM}{Long Short-Term Memory}
\newacronym{fmcw}{FMCW}{Frequency-Modulated Continuous Wave}

\begin{document}

\title{TinyRadarNN: Combining Spatial and Temporal Convolutional Neural Networks for Embedded Gesture Recognition with Short Range Radars}

\author{\IEEEauthorblockN{Moritz Scherer, Michele Magno, Jonas Erb, Philipp Mayer, Manuel Eggimann,  Luca Benini}
\thanks{M. Scherer, P. Mayer, M. Eggimann and L. Benini are with the Dept. of Information Technology and Electrical Engineering, ETH Z\"{u}rich, Switzerland (e-mail: 
\{scheremo, magnom, mayerph, meggiman, lbenini\}@iis.ee.ethz.ch).} 
\thanks{M. Magno is with the Dept. of Project-based Learning, ETH Z\"{u}rich, Switzerland (e-mail:
magnom@pbl.ee.ethz.ch).}
\thanks{J. Erb was with the Dept. of Information Technology and Electrical Engineering, ETH Z\"{u}rich, Switzerland at the time of his work on the TinyRadar project.}
\thanks{L. Benini is also with the Dept. of Electrical, Electronic and Information Engineering, University of Bologna, Italy.}
}

\markboth{
Accepted for publication at IEEE Internet of Things Journal}{
M. Scherer \MakeLowercase{\textit{et al.}}: TinyRadarNN: Combining Spatial and Temporal Convolutional Neural Networks for Embedded Gesture Recognition with Short Range Radars}

\IEEEoverridecommandlockouts
\IEEEpubid{\makebox[\columnwidth]{\copyright2020 IEEE \hfill} \hspace{\columnsep}\makebox[\columnwidth]{ }}

\maketitle

\begin{abstract}
This work proposes a low-power high-accuracy embedded hand-gesture recognition algorithm targeting battery-operated wearable devices using low power short-range RADAR sensors. A 2D Convolutional Neural Network (CNN) using range frequency Doppler features is combined with a Temporal Convolutional Neural Network (TCN) for time sequence prediction. The final algorithm has a model size of only 46 thousand parameters, yielding a memory footprint of only \SI{92}{KB}. Two datasets containing 11 challenging hand gestures performed by 26 different people have been recorded containing a total of 20'210 gesture instances. On the 11 hand gesture dataset, accuracies of 86.6\% (26 users) and 92.4\% (single user) have been achieved, which are comparable to the state-of-the-art, which achieves 87\% (10 users) and 94\% (single user), while using a TCN-based network that is 7500$\times$ smaller than the state-of-the-art. Furthermore, the gesture recognition classifier has been implemented on a Parallel Ultra-Low Power Processor, demonstrating that real-time prediction is feasible with only \SI{21}{\milli\watt} of power consumption for the full TCN sequence prediction network, while a system-level power consumption of less than \SI{120}{\milli\watt} is achieved. We provide open-source access to example code and all data collected and used in this work on tinyradar.ethz.ch.


\end{abstract}

\begin{IEEEkeywords}
gesture recognition, machine learning, internet of things, ultra-low power
\end{IEEEkeywords}
\section{Introduction}
\gls{hci} \revA{and \gls{har}} systems provide a plethora of attractive application scenarios with a wide array of solutions, strategies, and technologies \revA{\cite{naujoks2017human, Li2020}}. Hand gestures are one of the most natural ways for people to interact, control and engage with devices and machines in the \gls{iot} paradigm \cite{zhang2020learning}. For this reason it is not surprising that one of the emerging technologies in the context of wearable devices is gesture recognition \cite{wu2018orientation}. 
Traditionally, the approaches for capturing human gestures are based on image data or direct measurements of movement, i.e. by using motion sensors \cite{wu2018orientation,xu2011mems, wristflex}. The main types of sensors used in literature are cameras with \revA{and} without depth perception, force-sensitive resistors, capacitive elements \revA{and} accelerometers to measure the movement of the subject directly \cite{bandini2020analysis}. While these approaches have been shown to work well in controlled settings, robustness remains a challenge in real-world application scenarios. 
Image-based approaches have to deal with well-known environmental challenges like subject occlusion and variability in brightness, contrast, exposure and other parameters \cite{visionbased}. Another drawback of image-based solutions is the comparatively high power consumption, with commercial sensors like the Kinect sensors having power consumptions in the order of \revA{watts} \cite{kinectslides}. 
Wearable systems using motion-based sensing are much less affected by environmental variability and typically use significantly less power, but are more difficult to adapt to differences in user physique and behaviour. Approaches based on Wi-Fi have also been studied. On single subjects they have been shown to achieve high accuracy \cite{wristflex,tomo}, but are generally restricted to coarse or full-body gestures, due to the low spatial resolution, signal strength and susceptibility to electromagnetic interference and multi-path reflections \cite{8839094,9069270, 3dtracking}. 





A very promising, novel sensing technology for hand gesture recognition is based on high frequency and short-range pulsed RADAR sensors \cite{solipaper}. RADAR technology can leverage the advantages of image-based recognition with reduced challenges from environmental variability. The electromagnetic RADAR waves can propagate through matter, such that it can potentially record responses even if placed behind clothing. Furthermore, recently proposed designs based on novel sensor implementations can fit within a low power budget \cite{solipaper,eggimann2019low}, compatible with the constraints of wearable devices. However, achieving RADAR-based gesture recognition on the highly constrained computing platforms available on wearables remains an open challenge. 

Battery-operated wearable devices for the Internet of Things\revA{, especially those used for machine learning and data mining applications,} typically host an ARM Cortex-M or RISC-V based microcontroller, which can achieve power consumption in the order of a few \revA{milliwatts} and computational speeds in the order of hundreds of MOp/s \revA{\cite{Rahimi2016,Magno2017a, Savaglio2019, Fortino2014}}, while offering memory storage of at most a few \revA{megabytes}. Fitting within these limited computational resources to run machine learning algorithms especially for high-bandwidth sensors, such as imagers or RADARs, remains challenging \cite{chen2019deep, Samie2019}. 
Recently, several research efforts have started to focus on specialized hardware to run machine learning algorithms, and in particular neural networks on power-constrained devices \cite{9052677, chen2019deep,Cavigelli2018,Cavigelli2015a}. 
Parallel architectures leveraging near-threshold operation and multi-core clusters, enabling significant increases in energy efficiency, have been explored in recent years with different application workloads \cite{Gautschi2017} and low-power systems \cite{eggimann2019risc}.



The main state-of-the-art approaches to machine learning-based time-sequence modelling for gesture recognition are \gls{hmm} \cite{introhmm} and \gls{lstm} \cite{lstm} networks, which both use an internal state to model the temporal evolution of the signal. In recent years especially, \glspl{ann} have seen a rapid increase in popularity, with most recent works relying on \gls{lstm}-based approaches \cite{surveyhmm, surveyann}. 
On the processing side, previous work has shown the potential of RADAR signals for use with machine learning algorithms to classify static as well as dynamic hand gestures \cite{eggimann2019low}. The recently proposed RADAR sensing platform Soli, jointly developed by Infineon and Google, has been studied in different works, most prominently by Wang et al. \cite{interactingwithsoli}. They propose an \gls{lstm} model that achieves an accuracy of above 90\% over 11 classes. 

In contrast to the state-based modelling of the input signal, \glspl{tcn} are stateless in the sense that their computation model does not depend on the input. This means that they can compute sequential outputs in parallel, unlike \glspl{lstm} or \glspl{hmm} \cite{empiricalevaluation}. Furthermore, since they only use stateless layers, \glspl{tcn} use significantly less memory for buffering feature maps compared to \glspl{lstm}, defusing the memory bottleneck on embedded platforms. \glspl{tcn} have increasingly been adopted in many application scenarios where the classification of data is heavily linked to its temporal properties, for example, biomedical data \cite{jarrett2019dynamic} or audio data \cite{matthewdavies2019temporal}.

This paper proposes a novel embedded, highly accurate temporal convolutional neural network architecture, optimized for low-power microcontrollers. 
The proposed model achieves both a memory footprint of less than \SI{100}{KB}, as well as achieving a per-sequence inference accuracy of around 86.6\%  for 11 challenging gesture classes, trained on a multi-user dataset, and 92.4\% for a single-user dataset. This work exploits novel, low-power short-range A1 RADAR sensors from 
Acconeer\footnote{https://www.acconeer.com/products} to acquire two rich and diverse datasets, one for a single user and one for a total of 26 users, each containing 11 gestures. Further,  we leverage a multi-core RISC-V based embedded processor taking advantage of the emerging parallel ultra-low power (PULP) computing paradigm to enable the execution of complex algorithmic flows on power-constrained devices. \revA{Similar to Soli, possible deployment scenarios for the algorithm and processing platform include smart devices like smartphones\footnote{https://ai.googleblog.com/2020/03/soli-radar-based-perception-and.html} and smart thermostats\footnote{https://www.gearbrain.com/new-google-nest-hub-soli-2649744781.html} and even wearables with small form factor like smartwatches and hearing aids \cite{eggimann2019risc}. Due to the small footprint of less than \SI{30}{\milli\meter\squared}, the RADAR sensor can be easily integrated into most wearable devices \cite{Shahzad2021}.}

We show that highly-accurate, real-time hand gesture recognition within a power budget of around \SI{120}{\milli\watt}, including the sensor and processing consumption, is possible with the proposed sensor and computing platform. Experimental evaluations with a working prototype demonstrate both the power consumption and the high accuracy and are presented in the paper. 

The main contribution of this paper can be summarized as follows.

\begin{itemize}
    \item Design and implementation of a \gls{tcn} network architecture optimized for low-power hand gesture recognition on microcontrollers, achieving state-of-the-art accuracy with a total memory footprint of less than \SI{512}{KB}.
    \item \revA{Acquistion and labeling} of an open-source gesture recognition dataset featuring 11 challenging, fine-grained hand gestures recorded with the low-power Acconeer A1 pulsed RADAR sensor to provide a baseline dataset for future research.
    \item Implementation of the proposed model in a novel parallel RISC-V based microcontroller, \revA{featuring} 8 specialized parallel cores for processing and \SI{512}{KB} of on-chip memory. The novel, power-optimized architecture of the processors enables a full-system power consumption below \SI{100}{\milli\watt} in full active mode.  
    \item Evaluation of the benefits of the algorithm in terms of accuracy,  energy efficiency and inference speed, showing that the \revA{processor} consumes only \SI{21}{\milli\watt}, which is orders of magnitude less power for real-time prediction compared to the state-of-the-art, at a comparable level of accuracy.
\end{itemize}

\section{Related Work}
Hand gesture recognition is a widely investigated field. However, it is difficult to put all the research into context, as there are many different categories of hand gestures, which vary in complexity. Also, depending on the number of modelled gestures, the sensor used\revA{,} and how well diversified the studied dataset is, accuracies vary greatly.
In this section we review RADAR based approaches which are most directly comparable with our work.
We refer the interested reader to \cite{oneshotgesture, hmmgesture, hmmgesture2, 3dconv, lstmhmm, alldevices, Yang2019} for  image-based, inertial and RF-based gesture recognition.


\subsection{RADAR-based gesture recognition}
Some research has been conducted to exploit RADAR systems or radio signals to predict hand gestures. The approaches vary in terms of the application scenario, as well as accuracy and power efficiency. 
Different models without explicit sequence modelling have been employed in the past, a sample of which is discussed here.
Kim et al. use pulsed radio signals to determine static hand gestures by analysing the differences between reflected waveforms with the help of a 1D CNN. Accuracies of over 90\% are achieved for American Sign Language (ASL) hand signs using a \gls{cnn} and micro-Doppler signatures \cite{handgestrec2}. 
In their feasibility analysis, Kim and Toomajian use deep convolutional neural networks to classify ten hand-gestures using micro-doppler signatures from a pulsed RADAR. Their offline prediction algorithm reaches an accuracy of 85.6\% on a single participant \cite{handgestrec}.
Using a similar approach based on micro-doppler signatures and a \gls{fmcw} RADAR, Sun et al. showed that inference accuracy of over 90\% on a nine gesture dataset recorded from a stationary RADAR for driving-related gestures is possible \cite{sunann1}. 

Different works have used combinations of \gls{lstm} cells or Hidden Markov Models combined with different pre-processing strategies and convolutional layers to classify both coarse- and fine-grained gestures with the help of time-sequence modelling. 
Hazra et al. present a \gls{fmcw}-based system which is trained to recognize eight gestures, reaching an accuracy of over 94\% \cite{lstmcnn}.
\revA{Targeting} embedded, low-power applications, Lien et al. developed a high-frequency short-range RADAR specifically for the purpose of hand-gesture recognition, called \revA{Soli}. They implement a neural network to classify four hand gestures. Their final implementation uses a random forest classifier on those features with an optional bayesian filter of the random forest output. They use four micro-gestures, which they call "virtual button" (pinch index), "virtual slider" (sliding with index finger over thumb), "horizontal swipe" and "vertical swipe". On those four gestures, they achieve a per-sample accuracy of 78.22\% and a per-sequence accuracy of 92.10\% for the bayesian filtered random forest output \cite{solipaper}. 
Choi et al. used the \revA{Soli} sensor and a self-recorded 10 gesture dataset featuring ten participants to train an \gls{lstm}-based neural network. They achieve an accuracy of over 98\% using a GPU for inference computation \cite{lstm1}.  
Using the \revA{Soli} sensor, Wang et al. propose a machine learning model to infer the hand motions contained in the RADAR signal, based on an \gls{ann} network containing both convolutional layers and \gls{lstm} cells. They employ a fine-grained eleven gesture dataset recorded using the \revA{Soli} sensor.
While their approach shows a high average statistical accuracy of 87.17\%, their proposed model uses more than \SI{600}{MB} of memory which is several orders of magnitude more than most low-power microcontrollers offer. Moreover, the Soli sensors are consuming more than \SI{300}{\milli\watt} of power, which will drain any reasonably sized battery for a wearable device in a few minutes of use
\cite{interactingwithsoli}.

While it has been shown that \glspl{tcn} can outperform \glspl{lstm} for action segmentation tasks both in terms of accuracy and inference speed \cite{tcnpaper, empiricalevaluation}, the use of \glspl{tcn} for gesture recognition remains a relatively unexplored field of research. However, one work by Luo et al. indicates that classical 2D-\glspl{tcn} can perform equally well and even outperform approaches based on \gls{lstm} cells and \glspl{hmm} for gesture recognition tasks \cite{lstmtcn}.

This paper presents a combination of \gls{tcn} and \gls{cnn} models to improve energy efficiency, reduce memory requirements and maximize the accuracy of gesture recognition using sensor data from a short-range RADAR. The hardware implementation and the benefits of the combination of \gls{tcn} \revA{and \gls{cnn}} have briefly been discussed in the authors' previous work \cite{eggimann2019low}. 
In this paper, we significantly extend the contribution of the previous work by fully discussing the model architecture and comparing it against other state-of-the-art gesture recognition algorithms, showing that the proposed \gls{tcn}-based model performs significantly better in terms of accuracy per operation than the state-of-the-art \gls{lstm}-based approach.
We further evaluate in-depth the selection of features starting from the raw sensor data. We also show a full working implementation on an embedded platform and in-field measurements from a demonstrator.
To the best of the authors' knowledge, there is no previous work that evaluates the use of \glspl{tcn} for embedded, real-time hand-gesture recognition based on RADAR sensors.



\section{Background}
\subsection{Range Frequency Doppler Map}

\begin{figure}
  \begin{center}
    \includegraphics[width=\linewidth]{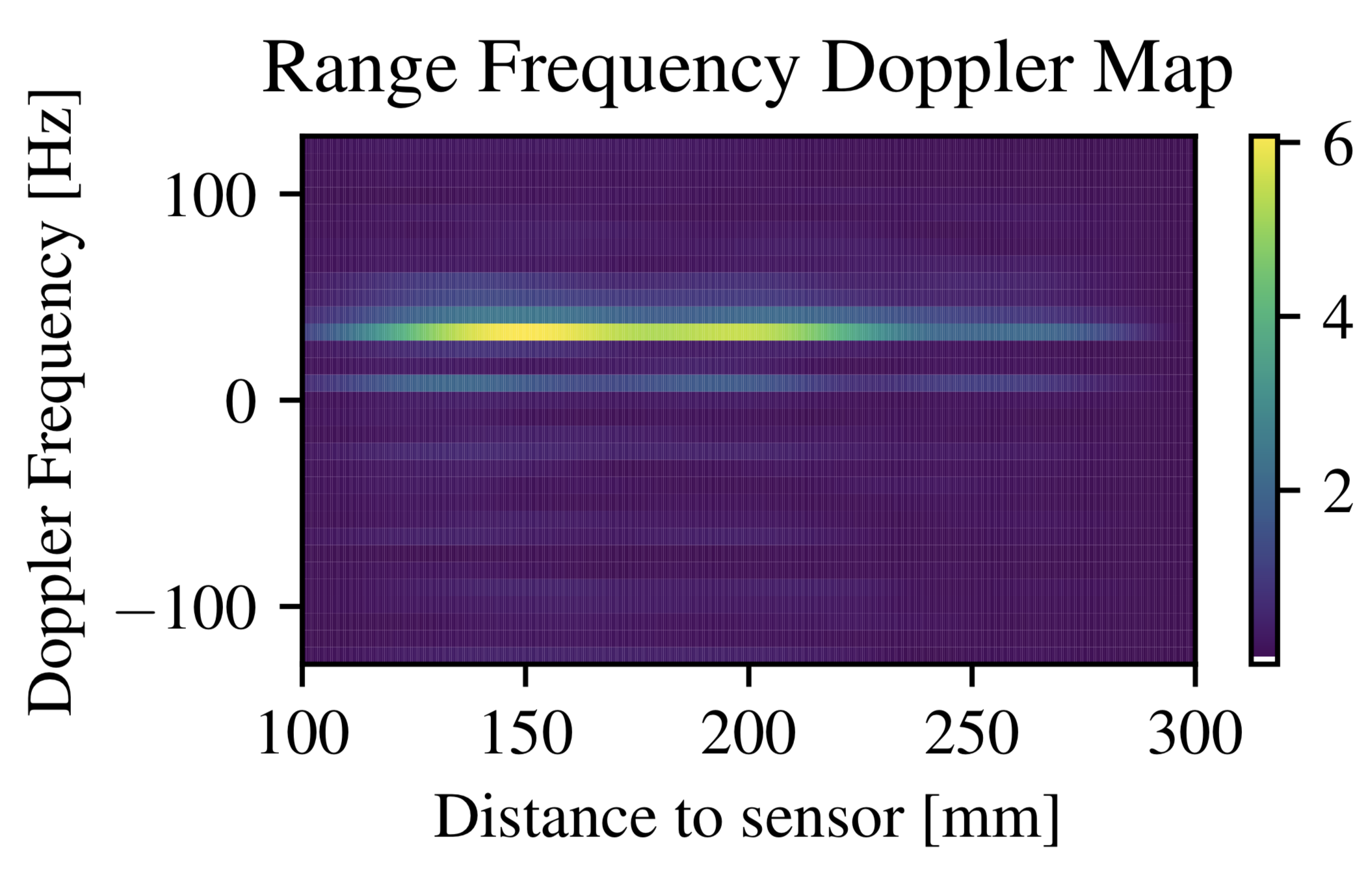}
    \caption{Range-Frequency Doppler Map for an example recording. The width and height dimensions correspond to the temporal frequency and the sampling range, respectively.}
    \label{fig:rfdm}
  \end{center}
\end{figure}

Feature maps based on the Fourier transform of the time axis, like the \gls{rfdm}, similarly to micro-Doppler signatures, have been proven to be effective for machine learning applications in previous research on gesture recognition \cite{handgestrec, sunann1, lstmcnn, lstm1, sunann2}.  It relies on the Doppler effect, which quantifies the shift of frequency in a signal that is reflected from a moving object. This shift of the frequency is correlated to the velocity of the object in the direction of the sensor. In order to detect changes in velocity, the I/Q signal is Fourier transformed into the frequency space, where changes in frequency can be observed. In order to detect the movement of objects in front of the sensor, multiple sweeps (i.e. time steps) are joined together and the time signal is Fourier transformed for each range point. As the sampled signal from each sweep $S(t,r)$ is time and range discrete the \gls{dft} is used. The transformed feature map $S(f,r)$ can be calculated according to the following equation:
$$S(f,r) = \sum_{t=0}^{T}S(t,r)e^{-\frac{2 \pi i f t}{T}}$$
Where $T$ is the total number of sample points per recorded distance point. In this work, only the absolute values of this function are considered.
An example \gls{rfdm} is shown in Figure \ref{fig:rfdm}.

\subsection{Temporal Convolutional Networks}

\begin{figure}
  \begin{center}
    \includegraphics[width=\linewidth]{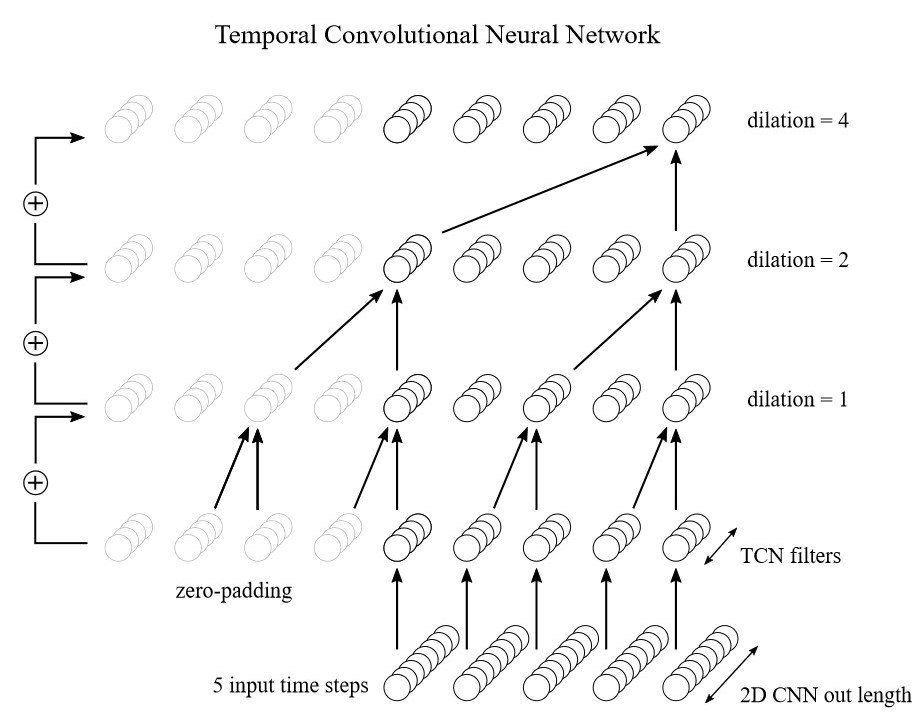}
    \caption{Layer structure of the TCN used in this work. Each input time step is one 1D vector that is generated by flattening the 2D CNN's output.
    The dilation factors used in the network are 1, 2 and 4. The kernel size for all convolutional filters in the TCN is 2.}
    \label{fig:TCNN}
  \end{center}
\end{figure}

Temporal Convolutional Networks are a modelling approach for time series using dilated 1D-convolutional neural networks, proposed by Lea et al. \cite{tcnpaper}, which has been used for a multitude of tasks, but very prominently in speech modelling \cite{wavenet, tcnn} and general human action recognition \cite{humanactiontcn}.
The basis of \glspl{tcn} are causal, dilated 1D-convolutions. Causal refers to the fact that for the prediction of any time step no future inputs are considered. Thus, the support pixel of the kernel is always chosen to be the last pixel. This is needed in a real-time prediction scenario, as in that case only the current and past data values are available at prediction time. 
To weigh past data for sequence predicition, \glspl{tcn} employ dilated convolutions over the temporal dimension. By increasing the dilation factor for consecutive layers the receptive field can be increased rapidly and very long effective memory of the network can be achieved. Figure \ref{fig:TCNN} shows the data flow of the \gls{tcn} as used in this work. The input data for the \gls{tcn} used in this work are the flattened, 1D outputs of a 2D-CNN.

Naturally, the \gls{tcn} produces one output per time step. In the following, we will refer to metrics considering each individual time step as \textit{per-frame} and to metrics considering the time step and all previous time steps modelled in the \gls{tcn} as \textit{per-sequence}.




\section{Low Power Short Range Radar and Dataset}

This chapter describes the properties of the Acconeer low power short-range RADAR sensor that was used in this work and the parameters of the datasets that were acquired using the sensor.

\subsection{Short Range RADAR for Gesture Recognition}

The RADAR devices used in this work are novel short-range pulsed \gls{radar} from Acconeer, pulsed with \SI{60}{GHz}. These low power devices use only one transmitter and receiver which reduces the power consumption to tens of Milliwatts.
The data returned by these sensors are sampled values of the I/Q signals. The \revA{RADAR sensor} is configured to continuously emit pulses at a fixed frequency of $f_{sweep}$, called \gls{rrf}. The time interval between two pulses is called \gls{rri}.

Let $t=0$ be the time at which the sensor sends out a pulse. Assuming that the transmitter and receiver are at the same position, i.e. being the same antenna, the response received at $t+2\Delta t$ corresponds to the reflection echo of an object located at a distance of $d=\frac{c}{2\Delta t}$ from the emitter/receiver, where $\Delta t$ is the time-of-flight of the pulse to the location of the object. \\
By regularly sampling the signal received after sending a pulse, a sweep vector containing reflections of objects at different distances can be computed. The distance resolution $\Delta d$ of the Acconeer sensor amounts to \SI{0.483}{\milli\meter}, which corresponds to a time-of-flight of \SI{1.6}{\nano\second}. 



\subsection{Dataset Specification and Acquistion}

\begin{figure}
  \begin{center}
    \includegraphics[width=\linewidth]{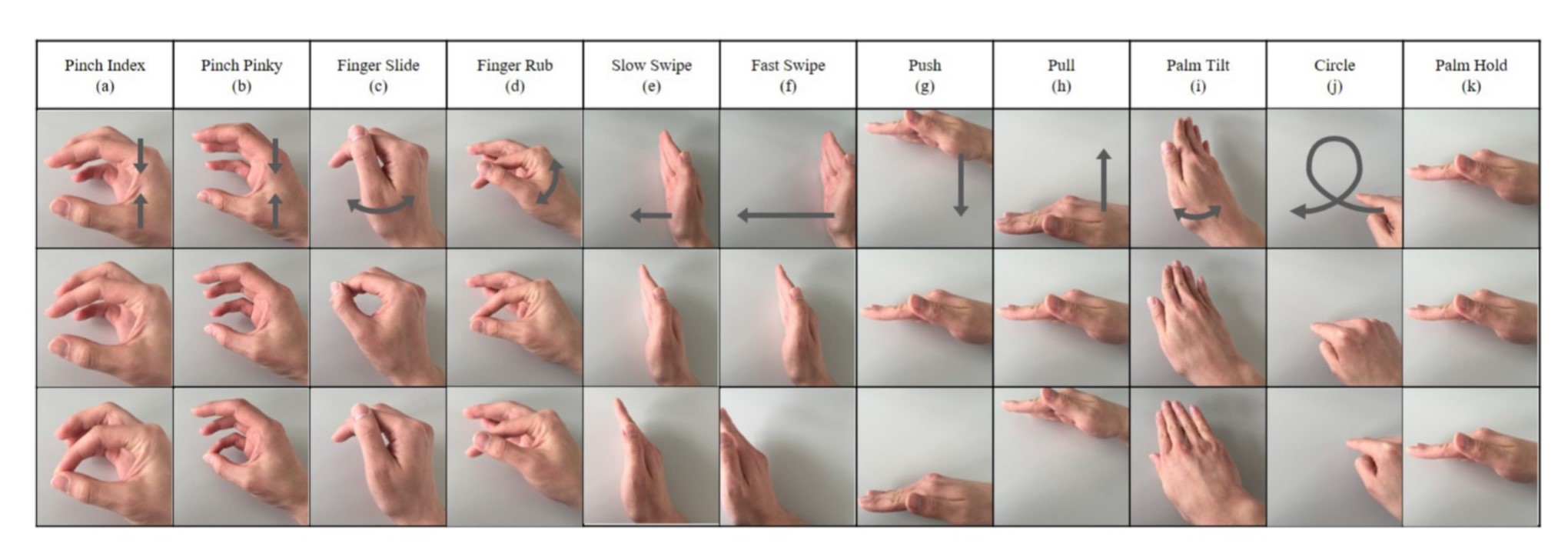}
    \caption{Overview of the gestures used in the dataset by Wang et al. \cite{interactingwithsoli} and this work. The eleven gestures contain fine-grained gestures like "Finger Slide", as well as coarser gestures like "Push" or "Pull".}
    \label{fig:gestures}
  \end{center}
\end{figure}

To train and evaluate the sensor for hand gesture recognition, two datasets were gathered in this work: One 5-gesture dataset and two 11-gesture datasets.\footnote{The 5G and 11G datasets and code for feature extraction are available for research purposes at https://tinyradar.ethz.ch} The 11-gesture data set features the same gestures as Wang et al. \cite{interactingwithsoli} and the 5-gesture dataset uses a subset of the same 11 gestures, consisting of the "Finger Slide", "Slow Swipe", "Push", "Pull" and "Palm Tilt" gestures. Using the same gestures as Wang et al. \cite{interactingwithsoli} allows for an effective comparison. All eleven gestures are depicted in \revA{Figure} \ref{fig:gestures}.

The 11-gesture dataset uses two Acconeer sensors with a sweep rate of \SI{160}{\hertz} each, while the 5-gesture dataset uses a single sensor with a sweep rate of \SI{256}{\hertz}.  Participants were shown \revA{Figure} \ref{fig:gestures}, the approximate height\revA{, \SI{20}{\centi\meter} above the sensor board,} at which to perform the gesture, but were given minimal instructions on how to perform the gestures. \revA{The gestures were performed in sitting position, without any additional inclination. The recording setup was not systematically varied between different persons and recordings.}
The 11-gesture dataset contains a total of 45 recording sessions of 26 different individuals, out of which 20 recordings are recorded from the same person to evaluate single-user accuracy, while the other 25 recordings are each recorded from different individuals. Subsets of the 11-gesture dataset are used to evaluate single user (SU) performance and multi-user (MU) performance. For the single-user dataset, the aforementioned 20 recordings from one single individual are used. For the multi-user dataset, one recording of the same individual is merged with the remaining 25 recordings of different individuals, which results in a dataset of 26 recordings of 26 different individuals. Thus, the multi-user and single-user datasets overlap by one recording of one individual.

A complete overview of the dataset parameters can be found in \revA{Table} \ref{tab:datasets}.

\begin{table}
    \centering
    \caption{Overview of the parameters used to record the dataset}
    \begin{tabularx}{\linewidth}{X|r r r}
         \textbf{Parameters} & \textbf{5-G} & \textbf{11-G (SU)} & \textbf{11-G (MU)}  \\
         \hline
         Sweep frequency & \SI{256}{\hertz} & \SI{160}{\hertz} & \SI{160}{\hertz} \\
         Sensors & 1 & 2 & 2 \\
         Gestures & 5 & 11 & 11 \\
         Recording length & \SI{3}{\second} & $\le$ \SI{3}{\second} & $\le$ \SI{3}{\second} \\
         \# of different people & 1 & 1 & 26 \\
         Instances per Session & 50 & 7 & 7 \\
         Sessions per recording & 10 & 5 & 5 \\
         Recordings & 1 & 20 & 26 \\
         Instances per gesture & 500 & 710 & 910 \\
         Instances per person & 2500 & 7700 & 35 \\
         Total Instances & 2500 & 7700 & 10010 \\
         Sweep ranges & 10 -- \SI{30}{\centi\meter} & 7 -- \SI{30}{\centi\meter} & 7 -- \SI{30}{\centi\meter} \\
         Sensor modules used & XR111 & XR112 & XR112
    \end{tabularx}
    \label{tab:datasets}
\end{table}

\section{Energy-Efficient and High Accuracy Gesture Recognition Algorithm}\label{chap:alg}


One of the major contributions of this paper is the proposal of a model to accurately classify hand gestures recorded with a short-range RADAR sensor. The proposed model enables the reduction of memory and computational resources, which pose the biggest challenge for the deployment of a model for small embedded devices such as microcontrollers.

The constraints for peak memory use and throughput in this work were chosen to work with microcontrollers like the ARM Cortex-M7 series and RISC-V based devices with a power budget in the order of tens of \revA{milliwatts}. These microprocessors are very memory-constrained, usually offering below \SI{512}{KB} of memory, and achieve optimal operating conditions when using 8-Bit quantization for the activations and 16- or 8-Bit quantization for the weights \cite{lai2018cmsis}. 

\subsection{Preprocessing}

Since the dataset consists of periodic samples of distance sweep vectors, we chose to use the well-known approach of stacking a number $TW$ of sweep vectors into one feature map window of raw data, which is called a frame. For the proposed network, the number of sweep vectors was chosen to be 32. This corresponds to a total time resolution of \SI{200}{\milli\second} per frame for 11-G datasets and \SI{125}{\milli\second} for the 5-G dataset.
These frames are then processed by normalizing them and computing their \gls{rfdm}. While the 2D range-frequency spectrum contains a real and an imaginary component, only the absolute value of each bin is used, since the phase component of the spectral representation, while having the same number of values as the magnitude, did not add any significant improvement to the overall inference accuracy. 

\subsection{Neural Network Design}

For the 11-G dataset, the input feature map size is $492 \times 32 \times 2$ values, as each sensor contributes one channel, the number of time steps considered are 32 and the number of range points per sweep is 492. Even when compressing each value to \SI{8}{bit}, the total required buffer memory for each frame amounts to \SI{246}{KB}. For successful time-sequence modelling, the information of multiple frames needs to be stored and processed. Using the raw frame for multiple time steps would lead to buffer space requirements in the order of megabytes, which is not available in commercial microcontrollers.

\begin{figure}
  \begin{center}
    \includegraphics[width=\linewidth]{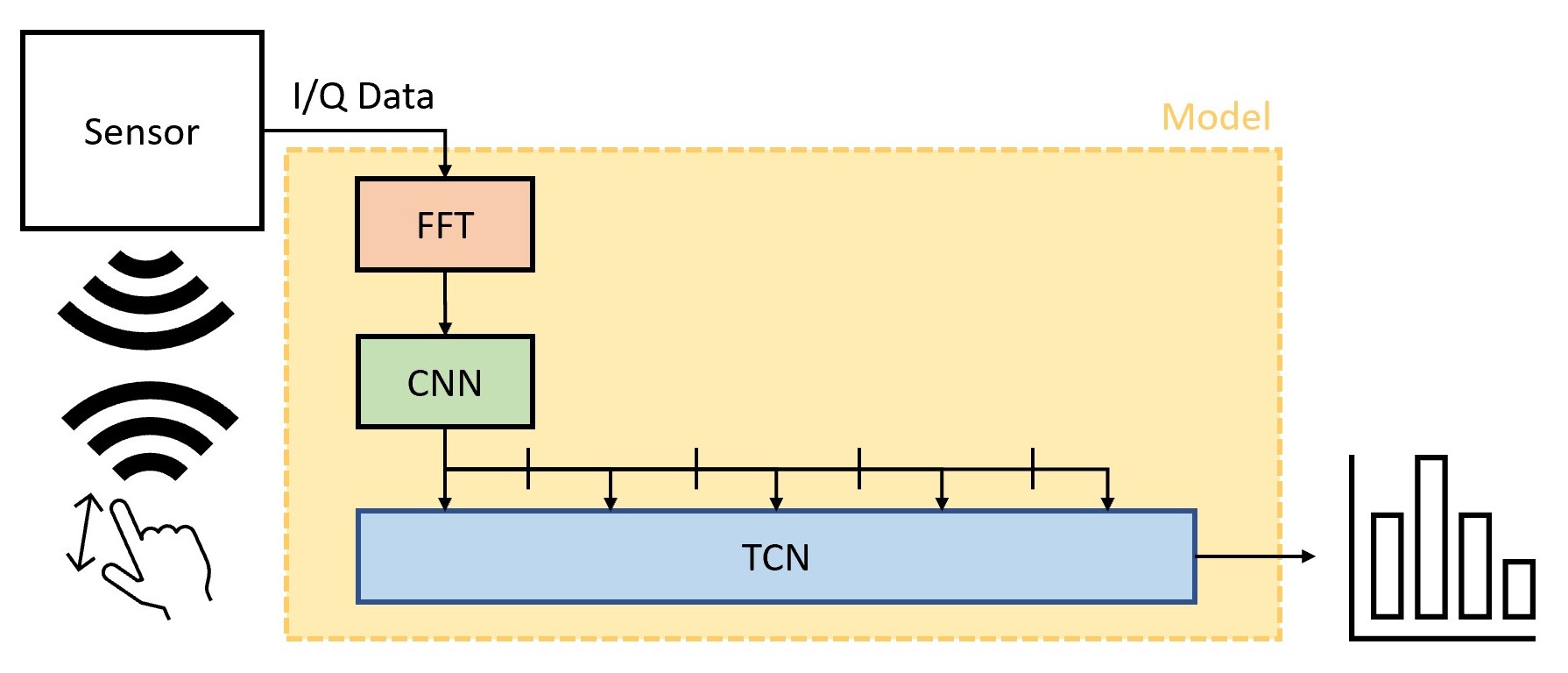}
    \caption{Overview of the processing algorithm. The raw I/Q sensor data is first processed by applying a Fourier transform, after which features are extracted from the frequency maps by processing them using a 2D CNN. The results of the feature extraction stage are flattened and five time steps are processed using a dilated TCN network.}
    \label{fig:overallalg}
  \end{center}
\end{figure}

To solve this issue, the proposed model is based on a combination of a 2D \gls{cnn} and a 1D \gls{tcn}, which are designed to separate the spatial-temporal modelling problem into two parts\revA{;} a short-term, spatial modelling problem, which captures little temporal information and can be solved on the level of individual frames, and a sequence modelling problem which can be solved on the level of extracted features from the first network. 
The overall data flow is depicted in Figure \ref{fig:overallalg}.

\subsection{Spatial and Short-Term Temporal Modelling}
\begin{figure}
  \begin{center}
    \includegraphics[width=\linewidth]{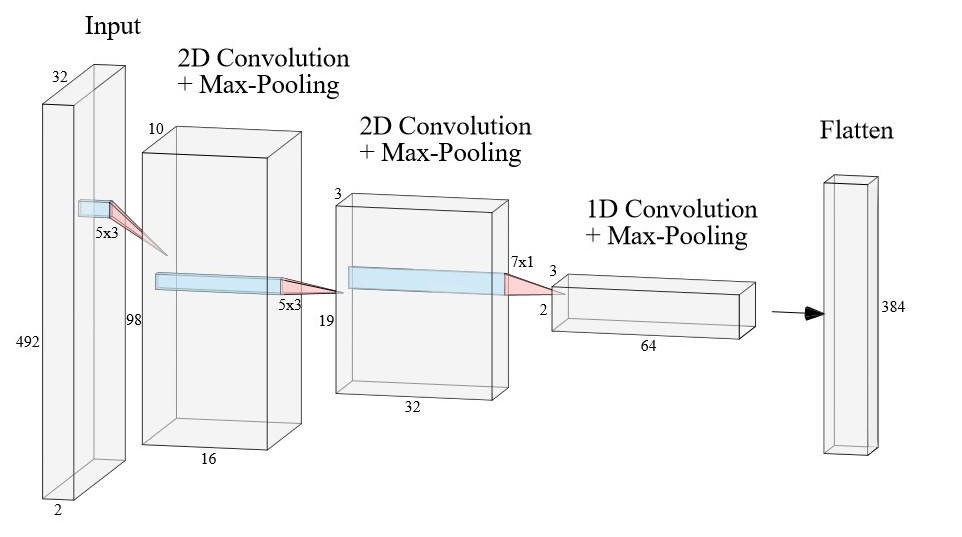}
    \caption{Layer structure of the 2D CNN. \revA{Each Convolutional layer is followed by a ReLU activation.}}
    \label{fig:cnn}
  \end{center}
\end{figure}

Spatial and short-term temporal modelling in this work can be seen as the task of extracting spatial and short-term temporal information from a single frame of RADAR data into a 1D feature vector containing spatial features that can be accurately classified with a sequence modelling algorithm. This approach compresses each frame by a factor of 82$\times$, which allows the extracted features to be stored on the low-memory microcontrollers for multiple time steps, which is required for accurate time-sequence prediction. The proposed network for spatial feature extraction is depicted in Figure \ref{fig:cnn}.

Since the width direction of the data frames corresponds to the spatial dimension, i.e. the distance from the sensor and the height direction corresponds to the temporal dimension of the frame, the frame width is considerably greater than the frame height. Since the distance sampling is chosen to be very fine-grained, wide kernels are used, both for pooling and convolutions. The layer parameters are shown in \revA{Table} \ref{tab:layerparams}. 
The total required buffer memory size for inference for algorithms using a static allocation of memory is given by the maximum of the sum of the buffer space required for the input and output feature map of any layer. For the proposed network, the total required buffer size is reached in the first layer and amounts to $(492 \cdot 32 \cdot 2 + 98 \cdot 10 \cdot 16) \cdot 8 \text{ Bit} = $ \SI{368}{KB}.

\begin{table}
    \centering
    \caption[Layer architecture of the 2D CNN]{Layer architecture of the 2D CNN}
    \begin{tabularx}{\linewidth}{X|r r r c}
    \textbf{Layer} & \textbf{Input} & \textbf{Output} & \textbf{Kernel} & \textbf{Padding} \\
    \hline
        2D Conv & 32$\times$492$\times$2 & 32$\times$492$\times$16 & 3$\times$5 & Same \\
        Max Pooling & 32$\times$492$\times$16 & 10$\times$98$\times$16 & 3$\times$5 & Valid \\
        2D Conv & 10$\times$98$\times$16 & 10$\times$98$\times$32 & 3$\times$5 & Same \\
        Max Pooling & 10$\times$98$\times$32 & 3$\times$19$\times$32 & 3$\times$5 & Valid \\
        1D Conv & 3$\times$19$\times$32 & 3$\times$19$\times$64 & 1$\times$7 & Same \\
        Max Pooling & 3$\times$19$\times$64 & 3$\times$2$\times$64 & 1$\times$7 & Valid \\
        Flatten & 3$\times$2$\times$64 & 384 & - & - \\

    \end{tabularx}
    \label{tab:layerparams}
\end{table}

\subsection{Long-Term Temporal Modelling}\label{chap:tcn}

The features computed by the 2D CNN are processed further with a \gls{tcn}. The \gls{tcn} uses an exponentially increasing dilation factor to combine features from different time steps into a single feature vector which can then be passed to a classifier consisting of fully-connected layers. For the proposed network, five time steps are considered by the \gls{tcn}, i.e. five consecutive output feature vectors of the 2D CNN are used as the input of the \gls{tcn}. This corresponds to a total effective time window of \SI{1}{s} for the 11-G datasets and \SI{.625}{\second} for the 5-G dataset.
The overall \gls{tcn} structure, taking into account the exponential dilation steps, is depicted in Figure \ref{fig:TCNN}.

\begin{figure}
    \centering
    \begin{minipage}{0.45\linewidth}
    \centering
        \includegraphics[width=0.9\linewidth]{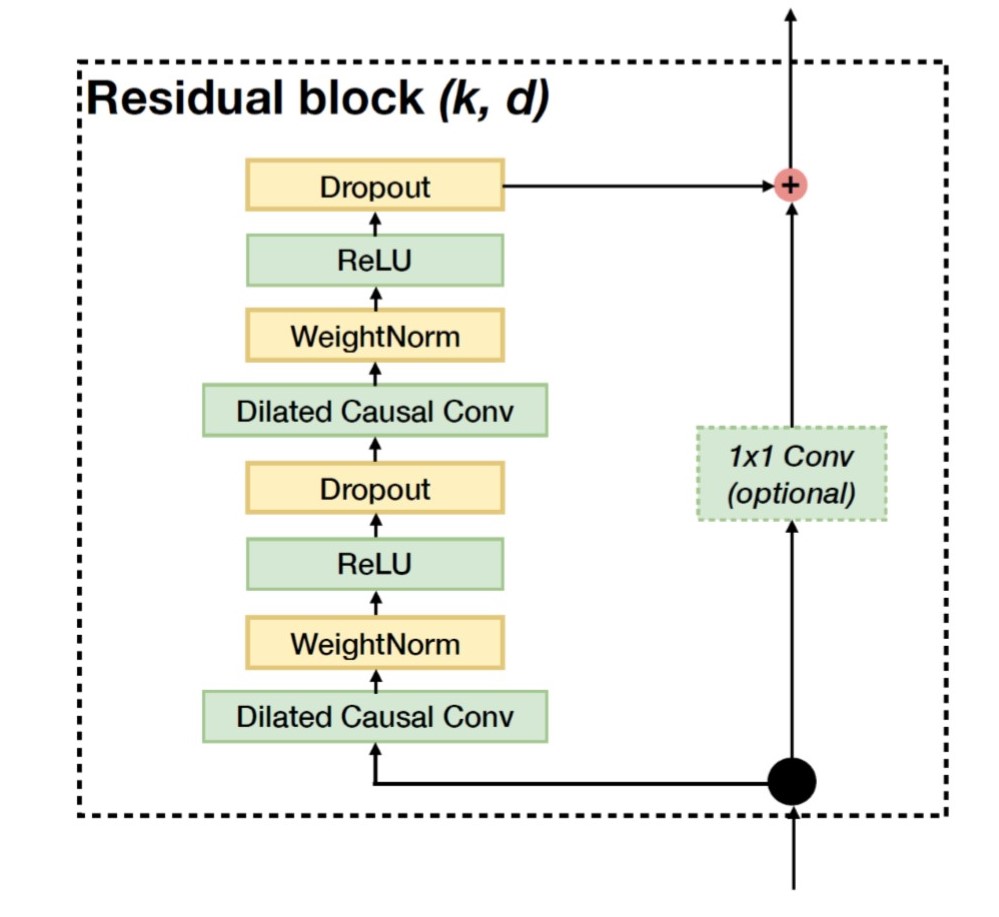}
        \label{fig:tcnog}
    \end{minipage}%
    \begin{minipage}{0.45\linewidth}
    \centering
        \includegraphics[width=0.9\linewidth]{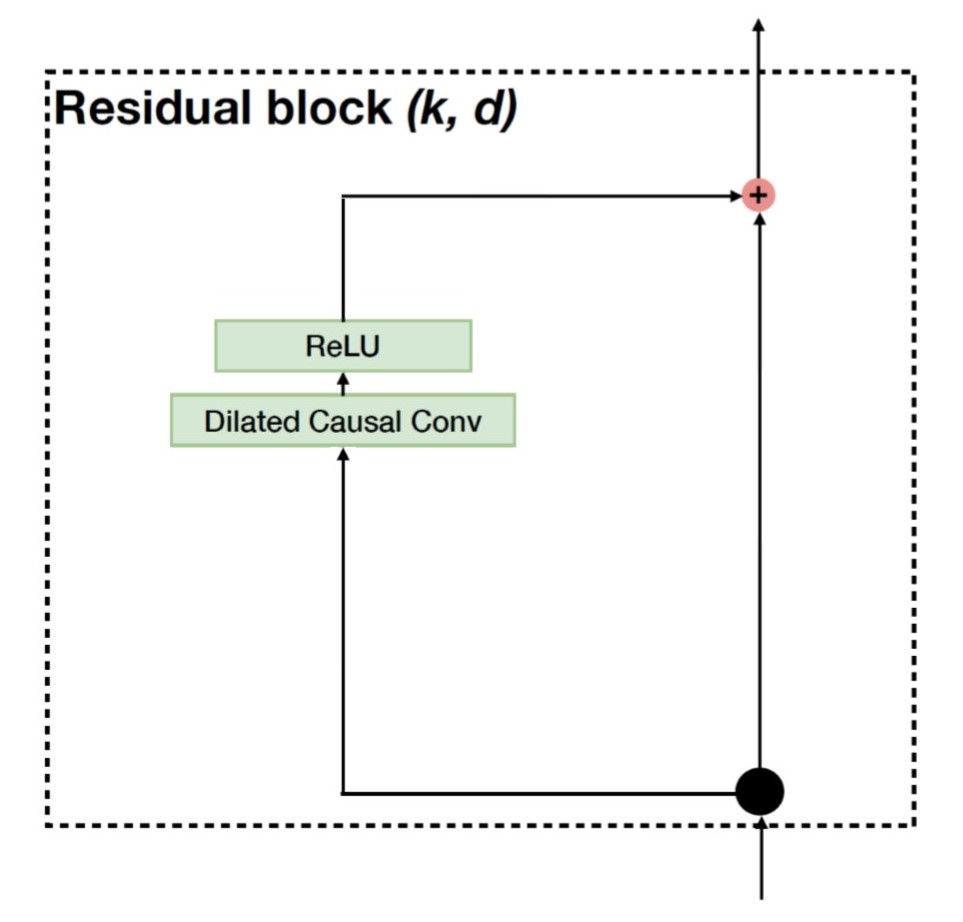}
        \label{fig:tcncust}
    \end{minipage}
    \caption{Comparison of the TCN residual blocks. The proposed network blocks (right) require a factor 2$\times$ less computations and memory than the original blocks (left), due to using only one convolutional layer instead of 2.}
\end{figure}

In this work, each \gls{tcn} filter in the \gls{tcn} \revA{consists} of residual blocks, each consisting of one depthwise convolution layer followed by a ReLU \cite{relu} activation, the result of which is then added to the original input. This is slightly different from the original definition of residual blocks in Lea et al. \cite{tcnpaper}, as normalization layers, dropout layers and one depthwise convolutional layer are removed to save memory space and execution time. A graphical comparison of the residual blocks as proposed by Lea et al. and as used in this work can be seen in Figures \ref{fig:tcnog} and \ref{fig:tcncust}. 

To reduce dimensionality, the output of the 2D CNN is filtered with a 1D Convolution which compresses the number of channels by a factor of 12$\times$. The compressed features are then collected for a total of five time steps before being passed to the dilated network. For the final output classification, the output of the dilated network is passed to three fully-connected layers.
The resulting network structure is described in \revA{Table} \ref{tab:tcnetwork}.

\begin{table}
    \centering
    \caption[\revA{Layer architecture of the TCN}]{Layer architecture of the TCN}
    \begin{tabularx}{\linewidth}{X|r r r r}
    \textbf{Layer} & \textbf{Input} & \textbf{Output} & \textbf{Kernel} & \textbf{Dilation} \\
    \hline
        Causal 1D Convolution & 5$\times$384 & 5$\times$32 & 1 & - \\
        Causal 1D Convolution & 5$\times$32 & 5$\times$32 & 2 & 1 \\
        Adding Layer & 5$\times$32 & 5$\times$32 & - & - \\
        Causal 1D Convolution & 5$\times$32 & 5$\times$32 & 2 & 2 \\
        Adding Layer & 5$\times$32 & 5$\times$32 & - & - \\        
        Causal 1D Convolution & 5$\times$32 & 5$\times$32 & 2 & 4 \\
        Adding Layer & 5$\times$32 & 5$\times$32 & - & - \\
        Fully connected & 5$\times$32 & 5$\times$64 & - & - \\
        Fully connected & 5$\times$64 & 5$\times$32 & - & - \\
        Fully connected & 5$\times$32 & 5$\times$11 & - & - \\
    \end{tabularx}
    \label{tab:tcnetwork}
\end{table}

\subsection{Training Setup}

Both the 2D CNN as well as the \gls{tcn} were implemented using the Keras/Tensorflow framework. The \gls{rfdm} features were extracted from the dataset and saved before training.
Both network parts were trained together, using a batch size of 128 for a total of 100 epochs. The optimizer chosen for training is Adam \cite{adam}. Both 5-fold cross-validation (CV5) and leave-one-user-out cross-validation (LOOCV) training runs were performed and are shown in the results section (Section \ref{sec:results}).

\section{Results and Discussion}\label{sec:results}

We evaluated the \revA{proposed} model and its implementation on embedded hardware in terms of power consumption and inference performance on \revA{the} system-scale. In particular, we present the test setup and the evaluation of the proposed model in terms of accuracy, memory and computational requirements in the first subsections, comparing different features and processing alternatives, while we present an evaluation of the implementation on a novel RISC-V-based parallel processor in a later subsection. 

\subsection{Experimental Setup}

The GAP8 from Greenwaves Technologies\footnote{https://greenwaves-technologies.com/ai\_processor\_gap8/} is an off-the-shelf RISC-V-based multicore embedded microcontroller developed for \gls{iot} applications. At its \revA{heart}, the GAP8 features one RISC-V microcontroller and an octa-core RISC-V processor cluster with support for specialized DSP instructions, derived from the PULP open-source project \cite{pulpcore}. The GAP8 memory architecture features two levels of on-chip memory hierarchy, containing \SI{512}{KB} of L2 memory and \SI{64}{KB} of L1 memory. 

\begin{figure}
  \begin{center}
    \includegraphics[width=\linewidth]{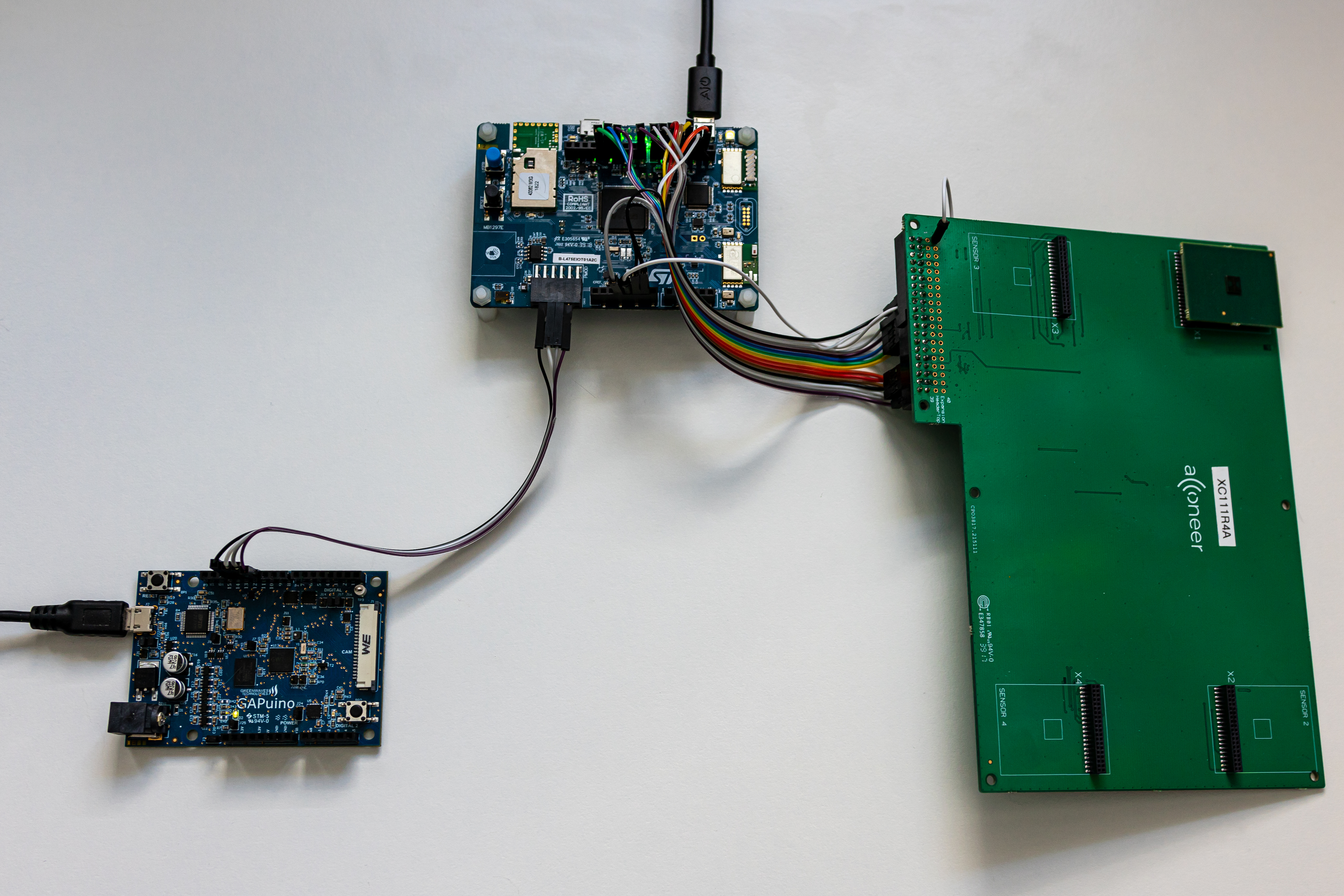}
    \caption{Picture of the hardware setup used to evaluate the system. The central board is a STM32L4 development board used to interface the RADAR sensor board (right) with the GAP8 development board (left).}
    \label{fig:hardwaresetup}
  \end{center}
\end{figure}

Figure \ref{fig:hardwaresetup} shows the hardware test setup, using evaluation boards for the GAP8 and A111 RADAR sensor, connected with an ARM Cortex-M4 evaluation board, which is used to broadcast the data to both a connected PC and the GAP8.

The trained model was deployed onto the GAP8 with the AutoTiler tool\footnote{\vspace{-2.6\baselineskip}
https://greenwaves-technologies.com/manuals/BUILD/AUTOTILER/html/index.html}, which generates C Code optimized for parallel execution of the model on the hardware platform.

\subsection{Accuracy of the Algorithm}
The inference accuracy of the algorithm can be discussed both in terms of per-frame accuracy, i.e. considering every frame for only one time step or in terms of per-sequence accuracy, i.e. the prediction for each frame taking into account the prediction for the individual frame at all time steps. 
To fairly compare results on the same dataset and frame definition, the per-frame metric is preferable, since it allows to accurately compare different approaches and the impact of sequence modelling versus single-frame processing.
For comparing to other datasets and frame definitions, the per-sequence accuracy is the preferable metric, since it levels out the impact of using frames with higher time resolution and represents more accurately how the network behaves in a practical setting.
The final results for the proposed network, both in terms of per-frame and per-sequence accuracy are shown in Table \ref{tab:res}.

\begin{table}
    \centering
    \caption[Per-frame and per-sequence inference accuracy of the full algorithm on the respective test/validation set]{Per-frame and per-sequence inference accuracy of the full algorithm on the respective test/validation set}
    \begin{tabularx}{\linewidth}{X|r r}
    \textbf{Metric} & \textbf{Per-Frame Accuracy} & \textbf{Per-Sequence Accuracy} \\
    \hline
    5-G SU-CV5 & 93.83\% & 95.00\% \\
    11-G SU-CV5 & 89.52\% & 92.39\% \\
    11-G MU-CV5 & 81.52\%  & 86.64\% \\
    11-G MU-LOOCV & 73.66\% & 78.85\% \\
    \end{tabularx}
    \label{tab:res}
\end{table}

For the following paragraphs, the per-frame accuracy is used to discuss the impact of changes in architecture and pre-processing, while the per-sequence accuracy is used to compare to other research. 

\subsection{Evaluation of Pre-Processing Methods}

To increase classification performance, different pre-extracted features were evaluated in combination with the features extracted by the convolutional neural network. The pre-extracted features are the signal energy, both for the \gls{sor} as well as the \gls{sot}, the signal variation for the \gls{sor} and \gls{sot} and the centre of mass, which measures the intensity of the signal over the range of the sensor.
An important consideration for embedded systems is the size of the feature maps since memory is the most common bottleneck for neural network implementations on microcontrollers and similar devices. An overview of the number of values per feature with respect to the number of sampling windows $TW$ and the number of range points $RP$ can be found in Table \ref{tab:features}.

\begin{table}
    \centering
    \caption[Overview of the size of different input features]{Overview of the size of different input features}
    \begin{tabularx}{\linewidth}{X|l r r}
    \textbf{Feature} & \textbf{Data Format} & \textbf{5-G} & \textbf{11-G} \\
    \hline
        Raw I/Q Signal & TW $\times$ RP $\times$ 2 & 26496 & 62976 \\
        Signal Variation 2D & (TW-1) $\times$ RP $\times$ 2 & 25668 & 61008 \\
        \gls{rfdm} & TW $\times$ RP & 13248 & 31488 \\
        Signal Energy \gls{sor} & RP & 414 & 492 \\
        Signal Energy \gls{sot} & TW & 32 & 32 \\
        Signal Variation \gls{sor} & RP & 414 & 492 \\
        Signal Variation \gls{sot} & TW & 32 & 32 \\
        Centre of mass & TW $\times$ 3 & 96 & 96
    \end{tabularx}
    \label{tab:features}
\end{table}

Due to the splitting of the data into windows containing both spatial and temporal information, an evaluation of the preprocessing and pre-extracted feature performance using the 2D CNN and a fully-connected layer to estimate the feature quality can be given. Using this setup, the per-frame training accuracy results in Table \ref{tab:featureacc} were achieved.

\begin{table}
    \centering
    \caption[Overview of the per-frame performance of different features for the 2D-CNN]{Overview of the per-frame performance of different features for the 2D-CNN}
    \begin{tabularx}{\linewidth}{X|r r}
    \textbf{Feature Combination} & \textbf{5-G SU-CV5} & \textbf{11-G MU-CV5} \\
    \hline
        Raw I/Q Signal & 90.35\% & 69.09\% \\
        Signal Variation 2D & 89.93\% & 65.32\% \\
        \textbf{\gls{rfdm}} & \textbf{91.08\%} & \textbf{69.37\%} \\
        Signal Energy \gls{sor} \&  \gls{sot} & 70.25\% & 51.90\% \\
        Signal Energy \gls{sor} & 65.67\% & 49.95\% \\
        Signal Energy \gls{sot} & 64.40\% & 40.72\% \\
        Signal Variation \gls{sor} & 38.10\% & 17.92\% \\
        Signal Variation \gls{sot} & 20.92\% & 10.57\%  \\
        Centre of mass & 47.56\% & 33.81\%
    \end{tabularx}
    \label{tab:featureacc}
\end{table}

The \gls{rfdm} features provide the best baseline in terms of pre-processed feature maps, both in terms of memory efficiency as well as classification performance. The raw data shows similar performance as the \gls{rfdm} in the case of a single-frame model, which makes it important to consider as using the raw data needs no pre-processing, while all other features do. However, the required energy to calculate the \gls{rfdm} features is around 34$\times$ less than what is used for one inference of the 2D-CNN, so the impact of pre-processing on energy efficiency is negligible.
To further increase the accuracy, combinations of the \gls{rfdm} with signal energy, variation and centre of mass were also studied. The per-frame performance of the \gls{rfdm} features combined with other features can be seen in Table \ref{tab:featurecomb}.

\begin{table}
    \centering
    \caption[Overview of the 2D-CNN per-frame network performance with combined features]{Overview of the 2D-CNN per-frame network performance with combined features}
    \begin{tabularx}{\linewidth}{X|r r}
    \textbf{Feature Combination} & \textbf{5-G SU-CV5} & \textbf{11-G MU-CV5} \\
    \hline
        \gls{rfdm} baseline & 91.08\% & 69.37\% \\
        \gls{rfdm} \& signal variation 2D & 91.05\% & \textbf{71.93\%} \\
        \gls{rfdm} \& signal energy \gls{sor} & 90.99\% & 70.24\% \\
        \gls{rfdm} \& signal variation \gls{sor} & 91.08\% & 69.16\% \\
        \gls{rfdm} \& centre of mass & \textbf{91.34\%} & 70.35\% \\
        \gls{rfdm} \& signal variation \gls{sot} & 76.93\% & 59.33\% \\
        \gls{rfdm} \& signal energy \gls{sot} & 91.20\% & 70.33\% \\
    \end{tabularx}
    \label{tab:featurecomb}
\end{table}

As already shown in the evaluation of pre-processing methods, the added features do not increase accuracy by a significant margin, which substantiates the choice not to add them for the proposed network.



\subsection{Hyperparameter Tuning of the TCN}

\begin{figure}
  \begin{center}
    \includegraphics[width=\linewidth]{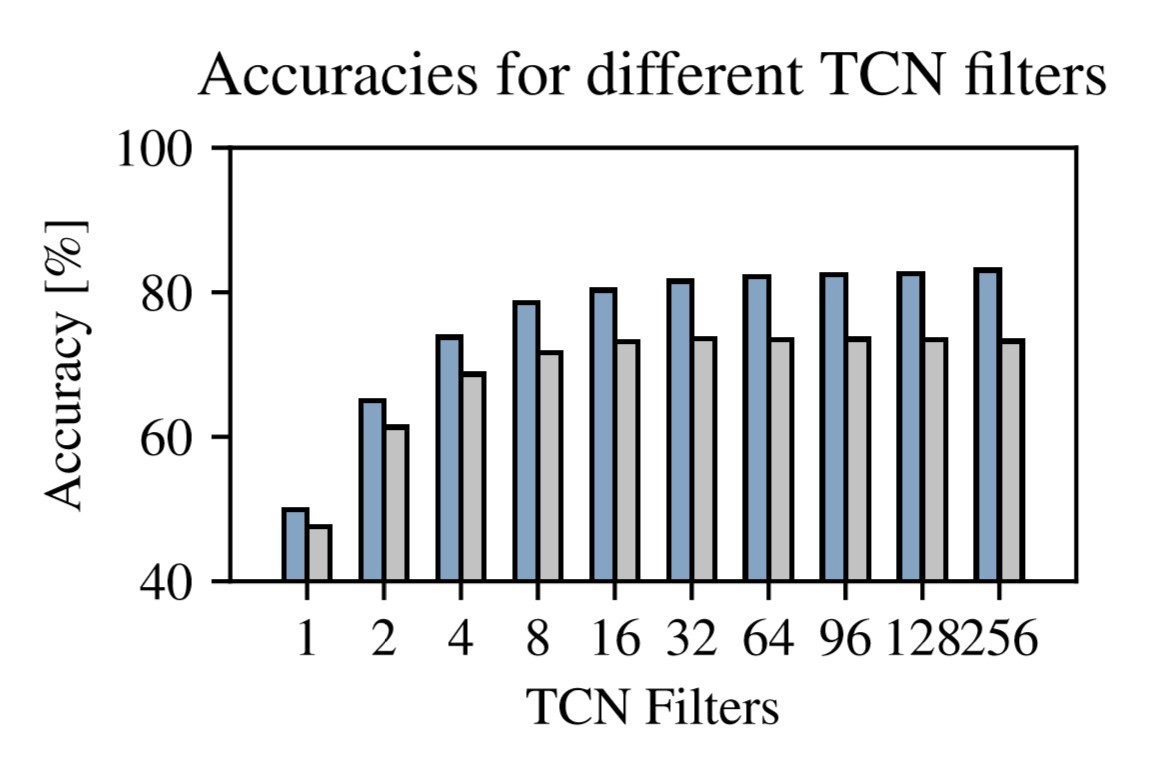}
    \caption{Classification performance vs. number of TCN filters on the 11-G dataset, using 5-fold cross-validation (blue) and leave-one-out cross-validation (grey). Even with exponential scaling of the number of filters, the accuracy stagnates after around 32 filters.}
    \label{fig:numfilt}
  \end{center}
\end{figure}

The performance of the network with the added \gls{tcn} was evaluated against the performance of the 2D CNN alone. As explained in section \ref{chap:tcn}, the number of \gls{tcn} filters is independent of the rest of the network and can be tuned to fit the constraints of the application and target hardware. To find the optimal operating point for the number of filters, the correlation between the number of filters and the increase in accuracy was evaluated for the 11 gesture dataset and is shown in Figure \ref{fig:numfilt}.

As can be seen in the graph, the classification accuracy plateaus after 32 \gls{tcn} filters. The averaged per-frame accuracy for different selections of features using 32 \gls{tcn} filters and five time steps can be seen in Table \ref{tab:finalperf}.

\begin{table}
    \centering
    \caption[Overview of the averaged per-frame accuracy of the whole network with combined features]{Overview of the averaged per-frame accuracy of the whole network with combined features}
    \begin{tabularx}{\linewidth}{X|r r}
    \textbf{Feature Combination} & \textbf{5-G SU-CV5} & \textbf{11-G MU-CV5} \\
    \hline
        Raw I/Q Signal & 91.90\% & 76.91\% \\
        \textbf{\gls{rfdm}} & \textbf{93.83\%} & \textbf{81.52\%} \\
        \gls{rfdm} \& signal variation 2D & 92.75\% & 78.84\% \\
        \gls{rfdm} \& signal energy \gls{sor} & 93.22\% & 80.92\% \\
        \gls{rfdm} \& centre of mass & 91.81\% & 78.45\% \\
        \gls{rfdm} \& signal energy \gls{sot} & 93.38\% & 78.99\% \\
    \end{tabularx}
    \label{tab:finalperf}
\end{table}

As previously discussed in the evaluation of the pre-processing methods, adding manually extracted features does not positively impact the overall accuracy of the network.

Further, for all combinations of features, especially with respect to the 11 gesture multi-user dataset, the \gls{tcn} improves the per-frame accuracy of the overall network by a significant margin. 

\subsection{Comparison to LSTM-based Networks}

The proposed model's time-sequence modelling network using custom \gls{tcn} layers was also evaluated against a modelling approach based on \gls{lstm}s as proposed by Schmidhuber et al. \cite{lstm} and a network using standard \gls{tcn} layers.

The performance for all three alternatives was evaluated using the same number of filters and time steps. The per-frame test accuracies for 32 and 128 filters are shown in Table \ref{tab:lstmvstcn32}.

\begin{table}
    \centering
    \caption[Per-frame test accuracy of the whole network for different sequence modelling approaches using 32 filters]{Per-frame test accuracy of the whole network for different sequence modelling approaches using 32 filters}
    \begin{tabularx}{\linewidth}{X|r r r}
    \textbf{Time steps} & \textbf{5} & \textbf{10} & \textbf{20} \\
    \hline
    LSTM, 32 filters & 79.24\% & 79.69\% & 80.71\%  \\
    LSTM, 128 filters & 79.29\% & 80.23\% & 81.77\%  \\
    Original TCN, 32 filters & 80.50\% & 80.46\% & 81.49\%  \\
    Original TCN, 128 filters & 80.55\% & 80.26\% & 82.09\%  \\
    Proposed TCN, 32 filters & 80.13\% & 80.17\% & 81.45\%  \\
    Proposed TCN, 128 filters & 80.79\% & 81.32\% & 82.79\%  \\
    \end{tabularx}
    \label{tab:lstmvstcn32}
\end{table}

The number of time steps beyond five does not significantly increase the inference performance of the network neither for the \gls{tcn} version nor for the \gls{lstm} version. Besides accuracy, the focus for embedded deployment is always on network size. Table \ref{tab:lstmvstcnparams} shows the number of parameters for 32 and 128 filters. Note that the number of time steps does not impact the number of parameters.

\begin{table}
    \centering
    \caption[Number of parameters required for sequence modelling using LSTM vs. TCN broken down by number of filters]{Number of parameters required for sequence modelling using LSTM vs. TCN broken down by number of filters}
    \begin{tabularx}{\linewidth}{X|r r r r}
    \textbf{Filters} & \textbf{32} & \textbf{64} & \textbf{96} & \textbf{128}  \\
    \hline
    LSTM & 25.4k & 99.8k & 223.5k & 396.3k \\
    Original TCN & 12.4k & 49.6k & 111.2k & 197.4k \\
    Proposed TCN & 6.2k & 24.8k & 55.6k & 98.7k \\
    \end{tabularx}
    \label{tab:lstmvstcnparams}
\end{table}

The number of parameters for the \gls{tcn}-based implementations is much lower than the number of parameters required for the \gls{lstm}-based implementations. Taking into account the superior accuracy, smaller memory footprint achieved with the \gls{tcn}-based implementations, the \gls{tcn} models perform better by all evaluated metrics. Furthermore, using the proposed \gls{tcn} variant, the number of parameters for the sequence modelling part can be reduced by a factor of 4$\times$ compared to LSTM-based variants.

\subsection{Experimental Results}

The proposed algorithm, as explained in section \ref{chap:alg}, was implemented and evaluated on a GAPuino evaluation board and power measurements were taken for both the microcontroller as well as the RADAR sensor. 
The overall number of weights of the model is split between the 2D CNN, requiring 22'368 weights and the TCN, requiring 22'917 weights. Using 16 bit quantization and considering the implementation overheads, the network requires just under \SI{92}{KB} on the GAP8. In terms of operations, the 2D CNN dominates the overall algorithm, taking up more than 99\% of the overall computations, which total around \SI{42}{MOps} per inference, taking a total of \SI{5.8}{MCycles} per inference on the GAP8.

\begin{figure}
  \begin{center}
    \includegraphics[width=\linewidth]{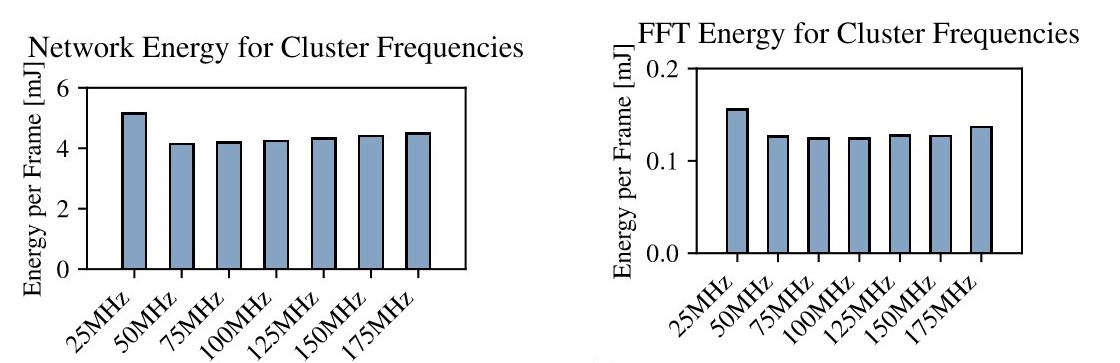}
    \caption{Overview of the microprocessor's energy efficiency while running the algorithm vs. its cluster frequency. To achieve real-time operation, at least 100 MHz are required.}
    \label{fig:energyvsfreq}
  \end{center}
\end{figure}

An overview of the energy consumption with respect to operating frequency is given in Figure \ref{fig:energyvsfreq}.

For the system to work in real-time at \SI{5}{\hertz} prediction rate, including the sampling of the RADAR sensor and execution of the algorithm, the cluster frequency should be chosen to be at least \SI{100}{\mega\hertz}. This leads to an average power consumption of \SI{21}{\milli\watt} of the GAP8 microcontroller measured during 2 inference/sleep cycles, with peak power consumption of \SI{98}{\milli\watt} while running the inference. An overall breakdown of operations, energy and cycles per inference at a clock frequency of \SI{100}{MHz} using 8 cores is shown in Table \ref{tab:breakdown}.

\begin{table}
    \centering
    \caption[Energy breakdown of the algorithm on GAP8 at 100 MHz]{Energy breakdown of the algorithm on GAP8 at 100 MHz}
    \begin{tabularx}{\linewidth}{X|r r r}
    \textbf{Algorithm step} & \textbf{Energy per Frame} & \textbf{Cycles} & \textbf{MACs} \\
    \hline
        FFT & \SI{0.12}{mJ} & $176 \cdot 10^{3}$ & - \\
        2D CNN & \SI{4.07}{mJ} & $5'100\cdot 10^{3}$ & $20'470\cdot 10^{3}$ \\
        TCN & \SI{0.32}{mJ} & $458\cdot 10^{3}$ & $256\cdot 10^{3}$ \\
        Dense & \SI{0.006}{mJ} & $86\cdot 10^{3}$ & $22\cdot 10^{3}$ \\
        \textbf{Full Network} & \SI{4.52}{mJ} & $5'820\cdot 10^{3}$ & $20'750\cdot 10^{3}$ \\
    \end{tabularx}
    \label{tab:breakdown}
\end{table}

To consider the overall system performance, the power consumption of the RADAR sensor has to be taken into account. Measuring the power consumption of the development board used in this work results in an upper bound, shown in Table \ref{tab:radarpower}.
\begin{table}
    \centering
    \caption[Power consumption of the RADAR sensor development board at different sweep frequencies]{Power consumption of the RADAR sensor development board at different sweep frequencies}
    \begin{tabularx}{\linewidth}{X|r r}
    \textbf{Sweep frequency} & \textbf{Power consumption} & \textbf{Samples} \\
    \hline
        \SI{100}{\hertz} & \SI{80}{\milli\watt} & 300 \\
        \SI{160}{\hertz} & \SI{95}{\milli\watt} & 480 \\
        \SI{256}{\hertz} & \SI{144}{\milli\watt} & 768 \\
    \end{tabularx}
    \label{tab:radarpower}
\end{table}

Taking into account the power consumption for the RADAR sensors, we arrive at a system-level power consumption of around \SI{200}{mW} when using two RADAR sensors at \SI{160}{Hz}, and \SI{115}{mW} when using one RADAR sensor at \SI{160}{Hz}.

\subsection{Comparison to Previous Work}

A direct comparison of this work is most directly possible with previous work in Wang et al. \cite{interactingwithsoli} since this work uses the same set of gestures and evaluation metrics. In Table \ref{tab:comparison} we compare our results with those reported by Wang et al. All accuracies are reported per-sequence, as the definition of frames is different in \cite{interactingwithsoli}.

\begin{table}
    \centering
    \caption[Comparison of the proposed implementation with previous work]{Comparison of the proposed implementation with previous work}
    \begin{tabularx}{\linewidth}{l|r r}
    \textbf{Metric} & \textbf{Interacting with Soli} & \textbf{This work} \\
    \hline
    Model size & \SI{689}{MB} & \SI{91}{KB} \\
    Single sensor power consumption & \SI{300}{mW} & \SI{95}{mW} \\ 
    Total sensor power consumption & \SI{300}{mW} & \SI{190}{mW} \\ 
    Network inference power & $ - $ & \SI{21}{mW} \\ 
    11-G SU Accuracy & 94.5\% & 92.39\% \\
    11-G MU-CV5 Accuracy & - & 86.64\%\\
    11-G MU-LOOCV Accuracy & 88.27\% & 78.85\% \\
    Number of different users & 10 & 26 \\
    \end{tabularx}
    \label{tab:comparison}
\end{table}

The direct comparison shows that our proposed network performs comparably accurately, if slightly worse, in all but leave-one-subject-out cross-validation, to the network proposed by \cite{interactingwithsoli}. Nonetheless, our network size is smaller by a factor of 7'500$\times$ and our power consumption is lower by several orders of magnitudes, as \cite{interactingwithsoli} use a GPU for inference, which operates at tens to hundreds of Watts of power consumption.
\section{Conclusion}
This work presented a high-accuracy and low-power hand-gesture recognition model combining a TCN and CNN model to achieve accuracy and low memory footprint. The model targets data processing with short-range RADAR. The paper proposed also a hand-gesture recognition system that uses low-power RADAR sensors from Acconeer combined with a GAP8 Parallel Ultra-Low-Power processor and can be battery operated. Two large datasets with 11 challenging hand-gestures performed by 26 different people containing a total of 20'210 gesture instances were recorded, on which the proposed algorithm reaches an accuracy of up to 92.4\%. The model size is only \SI{92}{kB} and the implementation in GAP8 shows that live-prediction is feasible with a power consumption of the prediction network of only \SI{21}{mW}. The results show the effectiveness and potential of RADAR-based hand-gesture recognition for embedded devices, as well as the network design, using the \gls{tcn} approach.
Further, we provide all necessary data and code to train the TinyRadarNN on tinyradar.ethz.ch. 

\section*{Acknowledgment}
\revA{The authors would like to thank \textit{armasuisse Science \& Technology} for funding this research.
The authors also thank Michael Rogenmoser and Cristian Cioflan for their valuable contributions to the research project.}

\bibliography{./main}

\begin{IEEEbiography}[{\includegraphics[width=1in,height=1.25in,clip,keepaspectratio]{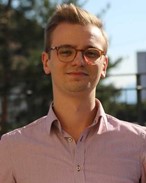}}]{Moritz Scherer} (GS’20) received the B.Sc. and M.Sc. degree in electrical engineering and information technology from ETH Zürich in 2018 and 2020, respectively, where he is currently pursuing a Ph.D. degree at the Integrated Systems Laboratory. His current research interests include the design of ultra-low power and energy-efficient circuits and accelerators as well as system-level and embedded design for machine learning and edge computing applications. He is the recipient of the Best Presentation Award at the 2019 IEEE Sensors and Applications Symposium. Moritz Scherer received the ETH Medal for his Master’s thesis in 2020.
\end{IEEEbiography}

\begin{IEEEbiography}[{\includegraphics[width=1in,height=1.25in,clip,keepaspectratio]{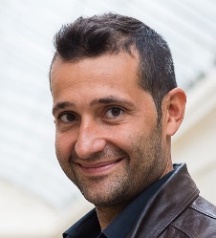}}]{Michele Magno} (SM’13) received his Masters and Ph.D. degrees in electronic engineering from the University of Bologna, Italy, in 2004 and 2010, respectively. Currently, he is a senior researcher at ETH Zurich, Switzerland, and Head of the Project-Based Learning Center at ETH Zurich. The key topics of his research are wireless sensor networks, wearable devices, machine learning at the edge, energy harvesting, power management techniques, and extended lifetime of batterie operated devices. He has collaborated with several universities and research centers, such as Mid University Sweden, where he is a guest full professor. He has published more than 150 papers in international journals and conferences, in which he got multiple best paper and best poster awards.

\end{IEEEbiography}

\begin{IEEEbiography}[{\includegraphics[width=1in,height=1.25in,clip,keepaspectratio]{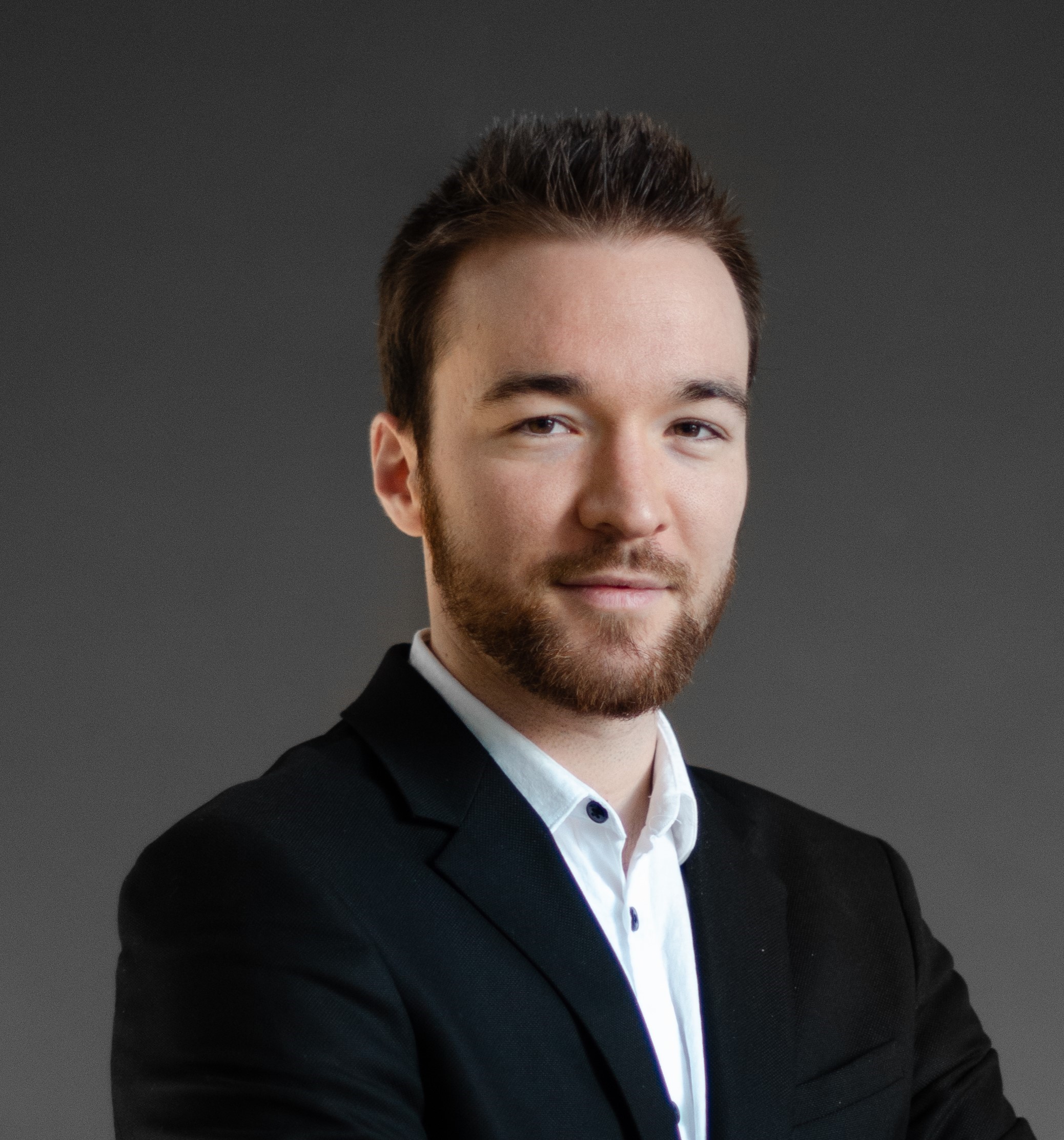}}]{Jonas Erb} received the B.Sc. and M.Sc. degree in electrical engineering and information technology from ETH Zurich in 2017 and 2019, respectively. After working in the Vast AG car rental start-up as an IT project manager in 2019, he started his own coaching business at the end of 2019 and works independently as an Embodied Authenticity coach since then, while following his passion for engineering in his spare time.
\end{IEEEbiography}

\begin{IEEEbiography}[{\includegraphics[width=1in,height=1.25in,clip,keepaspectratio]{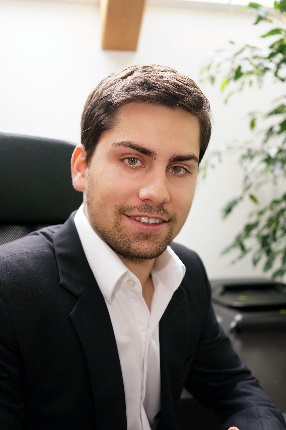}}]{Philipp Mayer} (GS’17) received his B.Sc. degree in electrical engineering and information technology from the TU Wien, Austria in 2016, and a consecutive M.Sc. degree from the ETH Zurich, Switzerland in 2018. He is currently pursuing a Ph.D. degree at the ETH Zurich Integrated System Laboratory. His research interests include low-power system design, energy harvesting, and edge computing. He is the recipient of the best paper award at the 2017 IEEE International Workshop on Advances in Sensors and Interfaces and the best student paper award at the 2018 IEEE Sensors Applications Symposium.
\end{IEEEbiography}

\begin{IEEEbiography}[{\includegraphics[width=1in,height=1.25in,clip,keepaspectratio]{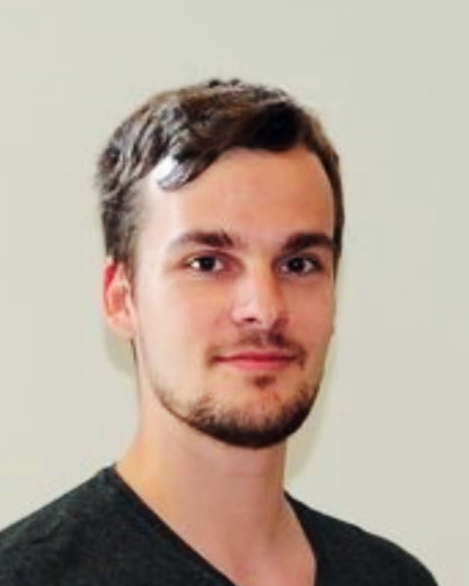}}]{Manuel Eggimann} (GS'18) Received his B.Sc. and consecutive M.Sc. degree in electrical engineering and information technology from the ETH Zurich, Switzerland in 2018.
He is currently pursuing a Ph.D. degree at the ETH Zurich Integrated Systems Laboratory.
His research interests include low-power hardware design, edge-computing and VLSI.
He is the recipient of the best paper award at the 2019 IEEE 8th International Workshop on Advances in Sensors and Interfaces.
\end{IEEEbiography}

\begin{IEEEbiography}[{\includegraphics[width=1in,height=1.25in,clip,keepaspectratio]{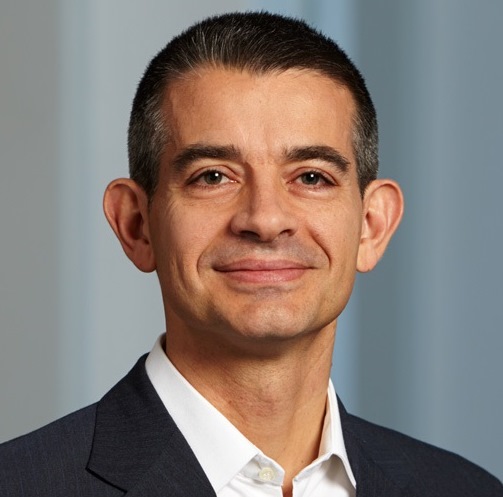}}]{Luca Benini} (F'07) is the Chair of Digital Circuits and Systems at ETH Zürich and a Full Professor at the University of Bologna. He has served as Chief Architect for the Platform2012 in STMicroelectronics, Grenoble. Dr. Benini’s research interests are in energy-efficient system and multi-core SoC design. He is also active in the area of energy-efficient smart sensors and sensor networks. He has published more than 1’000 papers in peer-reviewed international journals and conferences, four books and several book chapters. He is a Fellow of the ACM and of the IEEE and a member of the Academia Europaea.
\end{IEEEbiography}

\end{document}